\documentclass[12pt]{article}
\usepackage{amsfonts,amsmath,amssymb}
\usepackage{graphicx}
\usepackage{epic,eepic,color,graphicx}

\newfont{\twelvemsb}{msbm10 scaled\magstep1}
\newfont{\eightmsb}{msbm8}
\newfam\msbfam
\textfont\msbfam=\twelvemsb \scriptfont\msbfam=\eightmsb
\catcode`\@=11
\def\Bbb{\ifmmode\let\next\Bbb@\else
\def\next{\errmessage{Use \string\Bbb\space only in math mode}}\fi\next}
\def\Bbb@#1{{\fam\msbfam{{#1}}}}

\newcommand{\be}{\begin{equation}}
\newcommand{\ee}{\end{equation}}
\newcommand{\ba}{\begin{eqnarray}}
\newcommand{\ea}{\end{eqnarray}}

\begin{document}

\sloppy
\renewcommand{\thefootnote}{\fnsymbol{footnote}}
\newpage
\setcounter{page}{1} \vspace{0.7cm}
\begin{flushright}
20/01/09
\end{flushright}
\vspace*{1cm}
\begin{center}
{\bf Beyond cusp anomalous dimension from integrability }\\
\vspace{1.8cm} {Davide Fioravanti $^a$, Paolo Grinza $^b$,
Marco Rossi $^c$}\\
\vspace{.5cm} $^a$ {\em Sezione INFN di Bologna, Dipartimento di Fisica, Universit\`a di Bologna,} \\
{\em Via Irnerio 46, Bologna, Italy} \\
\vspace{.3cm} $^b${\em Departamento de Fisica de Particulas, Universidad de
Santiago de Compostela,} \\
{\em 15782 Santiago de Compostela, Spain} \\
\vspace{.3cm} $^c${\em Dipartimento di Fisica dell'Universit\`a della Calabria and
INFN, Gruppo collegato di Cosenza,}\\
{\em I-87036 Arcavacata di Rende, Cosenza, Italy}
\end{center}
\renewcommand{\thefootnote}{\arabic{footnote}}
\setcounter{footnote}{0}
\begin{abstract}
{\noindent We study the first sub-leading correction $O((\ln s)^0)$ to the cusp anomalous dimension  in the high spin expansion of finite twist operators in  ${\cal N}=4$ SYM theory. Since this approximation is still governed by a  linear integral equation (derived already from the Bethe Ansatz equations in a previous paper), we finalise it better in order to study the weak and strong coupling regimes. In fact, we emphasise how easily the weak coupling expansion can be obtained, confirms the known four loop result and predicts the higher orders. Eventually, we pay particular attention to the strong coupling regime showing agreement and predictions in comparison with string expansion; speculations on the 'universal' part (upon subtracting the collinear anomalous dimension) are brought forward.}
\end{abstract}
\vspace{6cm}

\newpage

\section{Plan of the work}
\setcounter{equation}{0}

Let us consider a (massless) gauge theory with a certain field $\phi$ in the adjoint
representation (or $q$ in the fundamental one). To fix ideas, we may focus our attention on the planar $sl(2)$ sector of
${\cal N}=4$ SYM with local operators
\begin{equation}
{\mbox {Tr}} ({\cal D}^s \phi^L)+.... \, , \label {sl2op}
\end{equation}
where ${\cal D}$ is the (symmetrised, traceless) covariant
derivative acting in all possible ways on the $L$ bosonic fields
$\phi$. At finite twist $L$ the high spin behaviour of the anomalous dimension
\begin{equation}
\gamma (g,s,L)=\Delta(g;s,L) -L-s \, , \label{gamma}
\end{equation}
will in general be determined by the universal (in the sense here of being twist and
flavour independent \cite{BGK, BES}, though theory dependent) scaling function $f(g)$
(cf. for instance \cite{Poly, KM, BGK, AMlong, FTT,BES} and references
therein):
\begin{equation}
\gamma (g,s,L)=f(g)\ln s + f_{sl}(g,L)+ \dots \, , \label{sud}
\end{equation}
(with a factor $1/2$ for the field $q$ in the fundamental representation
\cite{AMlong}) with this parametrisation of the 't Hooft coupling $\lambda = 8 \pi^2
g^2$. The universal scaling $f(g)$ equals in the $L=2$ sector twice the {\it cusp anomalous dimension} of the light-like Wilson loops \cite{KM} and therefore inherits this name.
We denote above by dots terms which are going to zero, whilst we wish to
investigate the sub-leading (constant) scaling $f_{sl}(g,L)$ (for QCD see also
\cite{MVV}). In fact, the leading scaling $f(g)$ has been constrained (for all the
values of $g$) in \cite{BES} (cf. also \cite{KL} for the relation with the Bethe
root density) by a linear integral equation derived from the {\it asymptotic} Bethe
Ansatz \cite{BSAF} and from the same Bethe Ansatz the sub-leading one $f_{sl}(g,L)$
has been shown in \cite{BFR}, after careful consideration \cite{BFRp}, to enjoy a
linear integral equation with the same kernel (but different inhomogeneous term; for
an alternative approach using the non-linear integral equation \cite{FMQR}, see \cite{FRS}).
It is, of course, non-universal in the sense again that it depends explicitly on the
twist $L$ (and, if there are other representations than the adjoint one, on the
representation carried by the field, as well as on the theory). Yet, it will be
conjectured in the following with some evidence to be wrapping-free (at least for
the smallest twist $L=2$), thus sharing the destiny of $f(g)$ in this respect  \footnote{As already clear, but also stressed in the following it is not universal as $f(g)$.}. The main evidence is
coming from the agreement with the recent string computations \cite{BFTT}, which
should seriously allow for gauge loop wrapping effects, as they live at very large
coupling $g$. In addition, a very recent analysis \cite {BJL} {\it $\grave {a}$ la} L\"uscher has proved for the scalar twist two operator that the wrapping correction at four loops goes to
zero (like $\ln^2s/s^2$).

In any case, thanks to Kotikov, Lipatov, Onishchenko and Velizhanin \cite{KLOV} it
is by now well established  that (in ${\cal N}=4$ SYM)
\begin{equation}
f_{sl}(g) \equiv f_{sl}(g,2)  \label{f_sl2}
\end{equation}
is the same for the operators with the smallest twist $L=2$ (and determines also
that of some twist three operators, see \cite{Becc3}). In fact, all twist two
operators belong to the same supermultiplet, and
their anomalous dimension is expressed in terms of a 'universal' function defined by
the scalars $\gamma_{univ.}(s)\equiv \gamma^\phi(s)$ and with shifted Lorentz spin
for gauginos $\gamma^\psi(s)=\gamma_{univ}(s+1)$ and gauge fields
$\gamma^A(s)=\gamma_{univ}(s+2)$.

Still at the leading twist $L=2$ Dixon, Magnea and Sterman \cite{DMS} have pointed
out that, although not universal (as $f(g)$), its {\it first logarithmic integral}, $f^{(-1)}_{sl}(g)$, equals the sub-leading correction to the logarithm of the (diverging) scattering amplitude (the following is just a sketchy formula whose precise definitions and regularisation rely on \cite{DMS} or references therein, in particular for the connexion between $f^{(-1)}_{sl}(g)$ and $f_{sl}(g)$ \footnote{The same connexion holds between $f^{(-1)}_{sl'}(g)$ and $f_{sl'}(g)$. A similar one between $f^{(-2)}(g)$ and $f(g)$ (cf. \cite{DMS} or references therein).})
\begin{equation}
\ln {\cal A}= \frac{f^{(-2)}(g)}{\epsilon^2} - \frac{f^{(-1)}_{sl'}(g)}{\epsilon} +.... \, ,
\label {ampli}
\end{equation}
in dimensional regularisation to $D=4-2\epsilon$, plus {\it the first logarithmic integral} of a universal function $h(g)$ (times a representation depending factor we omit here because not extant in $SYM_4$). In terms of the {\it by-logarithm-derivative} function of $f^{(-1)}_{sl'}(g)$, i.e. the so-called collinear anomalous dimension $f_{sl'}(g)$, this means the important relation between sub-leading term  
\begin{equation}
f_{sl}(g)= f_{sl'}(g)+ h(g) \, . 
\label {relation}
\end{equation}
In other words, this ($L=2$) sub-leading scaling in the anomalous dimension
(\ref{sud}) contains a universal part, $h(g)$, which, once subtracted from it,
yields (upon integration) the sub-leading correction in the amplitude (\ref{ampli}); both the
sub-leading terms are not universal, albeit their difference ought to be. As we will
see later (cf. (\ref{subleading})), the linear integral equation leads us to a split
form of $ f_{sl}(g,L)$ into a twist-dependent part and another independent of it,
$f_{sl}(g)$ (cf. (\ref{subleading}) below). The latter contains somehow the universal
$h(g)$ and thus might share with it, at very strong coupling, its logarithmic
behaviour (cf.  below (\ref{analytic}) and (\ref{fextra}) which are in agreement
with the string theory calculation by \cite{AM}). This may lead us to the possible
proposal
\begin{equation}
h (g) = f(g) \, [\ln \frac{1}{\sqrt 2 g} - (\frac{1}{2}-\frac{3}{2} \ln2) + \dots] \, ,
\end{equation}
with an intriguing object between square parenthesis (and the presence of $f(g)$ may
be perhaps motivated by the relation of $h(g)$ to a Wilson loop \cite{DMS}).
In fact, at the first order, recalling that $f(g)=2{\sqrt {2}} g+\dots$\, , we would
obtain
\begin{equation}
h (g) =  2{\sqrt {2}} g \,[ \ln \frac{1 }{\sqrt 2 g} - (\frac{1}{2}-\frac{3}{2}
\ln2) + \dots] \, ,
\end{equation}
which coincides with the tested function in string theory for gluons and quarks
\cite{Alday}.

{\bf Note added:} An interesting paper \cite {FZ}
appears today in the web archives. It seems to contain an analysis of the strong
coupling expansions of $f_{sl}(g)$ in overlap with some of the results
of this work.

\section{One loop results}
\setcounter{equation}{0}

The present analysis fully relies upon the main achievements of \cite {BFR},
where it was shown that the relevant integral equations which enter the
computation of the anomalous dimension at large spin are linear.
Let us briefly remind some results which will be useful later.

For what concerns the one
loop results, we consider equation (3.52) of that paper.
The meaning of such equation is that, in the high spin limit $s \rightarrow \infty$, the one loop
density of roots is approximated, up to orders $o(s^0)$, by the
function whose Fourier transform $\hat \sigma _0(k)$ reads as
\begin{equation}
\hat \sigma _0(k)=-2\pi \frac {\frac {L}{2}\left (1-e^{-\frac {|k|}{2}}
\right )+e^{-\frac {|k|}{2}}\left (1-
\cos \frac {ks}{\sqrt {2}}\right )}
{\sinh \frac {|k|}{2}}-4\pi \delta (k) \ln 2 +
o(s^0)
 \, . \label {s0}
\end{equation}
We also showed that the one loop energy,
\begin{equation}
E_0(s,L)=- \int _{-\infty}^{\infty} \frac {dk}{4\pi^2} \hat e(k) \hat
\sigma  _0 (k) - (L-2) e(0) \, , \nonumber
\end{equation}
with the function
\begin{equation}
e(u)=\frac {1}{u^2 +\frac {1}{4}} \Rightarrow \hat e (k)=2 \pi e^{-\frac {|k|}{2}} \, ,
\end{equation}
in the high spin limit behaves as
\begin{equation}
E_0(s,L)= 4 \ln s   - 4 (L-2) \ln 2 +  4 \gamma _E  +
o(s^0)  \label {E0} \, ,
\end{equation}
and the one loop anomalous dimension reads
\begin{equation}
\gamma_0(s,L)=g^2 E_0(s,L)= 4 g^2 \ln s   - [4 (L-2) \ln 2 -  4 \gamma _E] g^2  + g^2 o(s^0)  \label {gamma0} \, .
\end{equation}

\section{All loops results}
\setcounter{equation}{0}

In order to study the higher than one loop density of roots
$\hat \sigma _H (k)$ it is convenient to
define the quantity
\begin{equation}
 S(k)=\frac {2\sinh \frac {|k|}{2}}{2\pi |k|} \hat \sigma _H (k) \, .
\label{Sndefi}
\end{equation}
We then consider equation (4.11) of \cite {BFR}\footnote {The functions
$F_0(u)$, $F_H(u)$, appearing in (4.11) of \cite {BFR}, are
related to $\sigma _0(u)$ and $\sigma _H(u)$ by the relations
\begin {equation}
\sigma _0(u)=\frac {d}{du} F_0(u) \, , \quad \sigma _H(u)=\frac {d}{du} F_H(u) \, .
\end{equation}}.
Passing to Fourier transforms,
we obtain that the function $S(k)$ satisfies the linear equation
\begin{eqnarray}
&&S(k)=\frac {L}{|k|}[1-J_0({\sqrt {2}}gk)]
+  \frac {1}{\pi {|k|}} \int _{-\infty }^{+\infty}
\frac {dh}{|h|} \Bigl [ \sum _{r=1}^{\infty}
r (-1)^{r+1}J_r({\sqrt {2}}gk) J_r({\sqrt {2}}gh)\frac {1-{\mbox {sgn}}(kh)}{2}
e^{-\frac {|h|}{2}} + \nonumber \\
&+&{\mbox {sgn}} (h) \sum _{r=2}^{\infty}\sum _{\nu =0}^{\infty} c_{r,r+1+2\nu}
(g) \, (-1)^{r+\nu}e^{-\frac {|h|}{2}} \Bigl (
J_{r-1}({\sqrt {2}}gk) J_{r+2\nu}({\sqrt {2}}gh)-
J_{r-1}({\sqrt {2}}gh)
J_{r+2\nu}({\sqrt {2}}gk)\Bigr ) \Bigr ] \cdot \nonumber \\
&\cdot& \Bigl [ \frac {\pi |h|}{\sinh \frac {|h|}{2}}S(h)-
4\pi \ln 2 \ \delta (h)-\pi (L-2) \frac {1-e^{\frac {|h|}{2}}}{\sinh \frac {|h|}{2}}-2\pi \frac {1-
e^{-\frac {|h|}{2}}\cos \frac {hs}{{\sqrt {2}}}}{\sinh \frac {|h|}{2}} \Bigr ] +
o (s^0) \, , \label{Skeq}
\end{eqnarray}
where the functions $c_{r,s}(g)$ enter the definition of the "dressing factor"
(for their definition, see e.g. \cite {BES}).
As follows from the generalisation of the Kotikov-Lipatov identity \cite {KL},
the relation of the function $S(k)$ with the (all-loops) energy
(anomalous dimension) is
\begin{equation}
\gamma (g,s,L)=2 S(0) \, . \label {E-s}
\end{equation}
We now want to put equation (\ref {Skeq}) in a form which is suitable for
analysis of both the weak and the strong coupling limit.
It is convenient to restrict to the domain $k \geq 0$ and
expand $S(k)$ in series of Bessel functions
\begin{equation}
S(k)=\sum _{p=1}^{\infty} S_p(g) \frac {J_p({\sqrt {2}}gk)}{k} \,
\end{equation}
with the energy (\ref {E-s}) given by
\begin{equation}
\gamma(g,s,L)=\sqrt {2}g S_1(g) \, . \label {E-s1}
\end{equation}
On the other hand, for what concerns the coefficients $S_p(g)$,
after some simple calculations we find the following system of equations
\begin{eqnarray}
S_{2p-1}(g)&=&2 {\sqrt {2}}g (\ln s +\gamma _E) \delta _{p,1} +
4(2p-1)\int _{0}^{\infty}
\frac {dh}{h} \frac {\tilde J_{2p-1}({\sqrt {2}}gh)}{e^h-1} - \nonumber \\
&-& 2(2p-1)\, (L-2) \, \int _{0}^{\infty}
\frac {dh}{h} \frac {J_{2p-1}({\sqrt {2}}gh)}{e^{\frac {h}{2}}+1} - 2(2p-1)
\sum _{m=1}^{\infty} Z_{2p-1,m}(g) S_m (g) \, , \nonumber \\
\label {S0system} \\
S_{2p}(g)&=&4+ 8p \int _{0}^{\infty}
\frac {dh}{h} \frac {J_{2p}({\sqrt {2}}gh)}{e^h-1}
+2(L-2) - 4p \, (L-2)\, \int _{0}^{\infty}
\frac {dh}{h} \frac {J_{2p}({\sqrt {2}}gh)}{e^{\frac {h}{2}}+1} + \nonumber \\
&+&4p\sum _{m=1}^{\infty} Z_{2p,2m-1}(g) S_{2m-1}(g)
-4p\sum _{m=1}^{\infty} Z_{2p,2m}(g) S_{2m}(g) \, . \nonumber
\end{eqnarray}
In writing (\ref {S0system}) we used the notations
\begin{equation}
\tilde J_{2p-1}(x)=J_{2p-1}(x)-\delta _{p,1}\frac {x}{2} \, , \quad Z_{n,m}(g)= \int _{0}^{\infty}
\frac {dh}{h} \frac {J_{n}({\sqrt {2}}gh)J_{m}({\sqrt {2}}gh)}{e^h-1}
\, .
\end{equation}
The structure of the forcing terms appearing in the linear system
(\ref {S0system}) suggests that its solution and - consequently - the
(all-loops) anomalous dimension can be split as
\begin{equation}
S_r(g)=  S^{BES}_r(g) \, \ln s + (L-2) \, S^{(1)}_r(g)+ S^{extra}_r(g) \, ,
\label {splitting}
\end{equation}
where $S^{BES}_r(g)$ is the (coefficient of the) solution of the
BES equation \cite {BES}, $S^{(1)}_r(g)$ is the (coefficient of the) first
generalised scaling density \cite {FGR1,BK,FGR2,BF,FGR3} and the extra coefficients
$S^{extra}_r(g)$ are solutions of the following system
\begin{eqnarray}
S^{extra}_{2p-1}(g)&=&2 {\sqrt {2}}g \gamma _E \delta _{p,1} +
4(2p-1)\int _{0}^{\infty}
\frac {dh}{h} \frac {\tilde J_{2p-1}({\sqrt {2}}gh)}{e^h-1} - \nonumber \\
&-&  2(2p-1)
\sum _{m=1}^{\infty} Z_{2p-1,m}(g) S_m^{extra} (g) \, , \nonumber \\
\label {Sextsystem} \\
S^{extra}_{2p}(g)&=&4+ 8p \int _{0}^{\infty}
\frac {dh}{h} \frac {J_{2p}({\sqrt {2}}gh)}{e^h-1}
 + \nonumber \\
&+&4p\sum _{m=1}^{\infty} Z_{2p,2m-1}(g) S_{2m-1}^{extra}(g)
-4p\sum _{m=1}^{\infty} Z_{2p,2m}(g) S_{2m}^{extra}(g) \, . \nonumber
\end{eqnarray}
From the splitting (\ref {splitting}) of their solution,
we obtain that the all loops energy at high spin is
\begin{equation}
\gamma(g,s,L)=f(g) \ln s  + (L-2) f^{(1)} (g)+ f_{sl}(g) + o(s^0)\, ,
\label {Esplitting}
\end{equation}
where $f(g)$ is the cusp anomalous dimension, $f^{(1)} (g)$ is the first generalised scaling
function and $f_{sl}(g)=\sqrt {2}g S_1^{extra}(g)$ comes from the
solution of (\ref {Sextsystem}). In other terms, the sub-leading correction is worth
\begin{equation}
f_{sl}(g,L)= f_{sl}(g) + (L-2) f^{(1)} (g) \, ,
\label {subleading}
\end{equation}
with explicit mention to its twist dependence (since $f_{sl}(g)$ does not depend on $L$).

In the next sections we will analyse the weak and the strong coupling expansion of
$f_{sl}(g,L)$.

\section{Weak coupling}
\setcounter{equation}{0}

Since the system (\ref {Sextsystem}) is linear (as the original linear integral equation \cite{BFR}), it can be systematically and very easily expanded at small $g$, giving the weak coupling (convergent) series of  $f_{sl}(g,L)$ (via (\ref{subleading})). This kind of computations can be automatised by using a program for the symbolic manipulation in order to go on to an arbitrary order \footnote{In what follows we report the results up to six loops, but we easily reached loop ten. We invite the interested reader to contact the authors for having these expressions, since their writing would fill some pages in.} .

In view of (\ref{subleading}), it is natural to start by the twist two formula up to the order $g^{12}$:
\begin{eqnarray}
&&  f_{sl}(g)  =  \gamma_E \, f(g)-24 \zeta (3) \, \left( \frac{g}{\sqrt 2} \right)^4  +
\nonumber  \\
&& + \frac{16}{3} \left(\pi ^2 \zeta (3)+30 \zeta (5)\right) \left( \frac{g}{\sqrt 2} \right)^6 + \nonumber \\
&& -\frac{8}{15} \left(7 \pi ^4 \zeta (3)+50 \pi ^2 \zeta (5)+2625 \zeta (7)\right)  \left( \frac{g}{\sqrt 2} \right)^8 + \nonumber  \\
&&+ \left( \frac{128}{35} \pi ^6 \zeta (3)+192 \zeta (3)^3+\frac{832}{45} \pi ^4 \zeta (5)+\frac{560}{3} \pi ^2 \zeta (7)+14112 \zeta (9) \right) \left( \frac{g}{\sqrt 2} \right)^{10} +   \nonumber  \\
&&-\Bigg( \frac{58552 \pi ^8 \zeta (3)}{14175}+\frac{1184}{63} \pi ^6 \zeta (5)+\frac{5824}{45} \pi ^4 \zeta(7)+\frac{32}{3} \pi ^2 \left(10 \zeta (3)^3+147 \zeta (9)\right)+ \nonumber \\
&&+32 \left(164 \zeta (3)^2 \zeta (5)+4851 \zeta(11)\right)  \Bigg)
\left( \frac{g}{\sqrt 2} \right)^{12} + \ldots \, .
\end{eqnarray}
And we end up with the general twist expansion (we also expanded $ f^{(1)}(g)$ according to \cite{FRS, BFR})
\begin{eqnarray}
&& f_{sl}(g,L) = f_{sl}(g) + (L-2) f^{(1)}(g) =  (\gamma_E - (L-2) \ln 2)f(g)+
8 (2 L-7) \zeta (3) \, \left( \frac{g}{\sqrt 2} \right)^4  +
\nonumber  \\
&& -\frac{8}{3} \left(\pi ^2 \zeta (3) (L-4)+3 (21 L-62) \zeta (5)\right) \left( \frac{g}{\sqrt 2} \right)^6 + \nonumber \\
&& +\frac{8}{15} \left(\pi ^4 \zeta (3) (3 L-13)+75 (46 L-127) \zeta (7)+5 (11 L-32) \pi ^2 \zeta (5)\right) \left( \frac{g}{\sqrt 2} \right)^8 + \nonumber  \\
&& - \Bigg( \frac{128}{945} \pi ^6 \zeta (3) (11 L-49)+8 \left(2695 \zeta (9) L+16 \zeta (3)^3 L-7154 \zeta (9)-56 \zeta (3)^3\right) + \nonumber \\
&& +\frac{40}{3} (25 L-64)
   \pi ^2 \zeta (7)+\frac{8}{45} (103 L-310) \pi^4 \zeta (5) \Bigg) \left( \frac{g}{\sqrt 2} \right)^{10} +   \nonumber  \\
&&+ \Bigg( \frac{32}{45} \pi ^4 \zeta (7) (295 L-772)+\frac{8}{3} \pi ^2 \left(1519 \zeta (9) L+24 \zeta (3)^3 L-3626 \zeta (9)-88 \zeta (3)^3\right) + \nonumber \\
&& +8 \left(33285 \zeta (11) L+536 \zeta (3)^2 \zeta (5) L-85974 \zeta (11)-1728 \zeta (3)^2 \zeta (5)\right) + \nonumber \\
&& +\frac{8}{945} (2023 L-6266) \pi ^6 \zeta
   (5)+\frac{8 (2956 L-13231) \pi ^8 \zeta (3)}{14175} \Bigg) \left( \frac{g}{\sqrt 2} \right)^{12}+\ldots \, .
\end{eqnarray}
Importantly, this expansion for twist $L=2, 3$ agrees with those up to four loops present in the literature \cite{KLOV, B&KLRSV, FRS}  and checked by different means (cf., for instance, \cite{BFTT} for a summary). Thanks to \cite{BJL, BeFo} we can also argue that for twist two (at least up to four loops) $f_{sl}(g)$ enjoys the property of being wrapping free, and we would like to conjecture the same in general about $f_{sl}(g, L)$.

In fact, last but not least, after the completion of this work we became aware \footnote{We ought to thank Matteo Beccaria for this private communication.} that a five loop calculation (wrapping inclusive) on twist $L=3$ would exactly confirm the expansion of this section.

\section{Strong coupling}
\setcounter{equation}{0}

Let us pass to consider the strong coupling limit of
(\ref {Sextsystem}). From the results in Appendix A, we get that the leading contribution at large
$g$ in the forcing terms of (\ref {Sextsystem}) is
\begin{equation}
-2{\sqrt {2}} g \ln g \, \delta _{p,1} \, ,
\end{equation}
which allows to write down the following analytic leading behaviour
\begin{equation}
f_{sl}(g)=- f(g) \, \ln g+ ....
\label{analytic}
\end{equation}
where $f(g)$ is the cusp anomalous dimension at strong coupling \cite {BKK,KSV}
and the dots denotes contributions
of order $g$ or smaller. This important point will be clarified in the remaining part of the section.

In order to analyse the next to leading terms of $f_{sl}(g)$ at large $g$,
we solve numerically the linear system (\ref {Sextsystem}).
Indeed, this system is amenable for the usual numerical treatment as developed
in \cite{BBKS}. The kernel is the same as in the study of the
universal scaling function, settling any issue about convergence. What
changes now is the nature of the forcing terms, which is expected to produce some subtle effects at strong coupling.
In particular, as explained in appendix A, the forcing term which
contains $\tilde{J}_1(x)$ shows a logarithmic leading asymptotic behaviour
for $g \to \infty$.

A (numerically) clean way to study the system at large $g$ is to
explicitly subtract such a logarithmic dependence in equations
(\ref{Sextsystem}). We are then led to consider the objects
$\tilde S^{extra}_r(g)$, satisfying the linear system
\begin{eqnarray}
\tilde S^{extra}_{2p-1}(g)&=&2 {\sqrt {2}}g \left(\gamma_E \, - \, \ln \frac{2 \sqrt 2 }{g} \right)  \delta _{p,1} +
4(2p-1)\int _{0}^{\infty}
\frac {dh}{h} \frac {\tilde J_{2p-1}({\sqrt {2}}gh)}{e^h-1} - \nonumber \\
&-&   2(2p-1)
\sum _{m=1}^{\infty} Z_{2p-1,m}(g) \tilde S_m^{extra} (g) \, , \nonumber \\
\label {Sextsystemmodif} \\
\tilde S^{extra}_{2p}(g)&=&4+ 8p \int _{0}^{\infty}
\frac {dh}{h} \frac {J_{2p}({\sqrt {2}}gh)}{e^h-1}
 + \nonumber \\
&+&4p\sum _{m=1}^{\infty} Z_{2p,2m-1}(g) \tilde S_{2m-1}^{extra}(g)
-4p\sum _{m=1}^{\infty} Z_{2p,2m}(g) \tilde S_{2m}^{extra}(g) \, . \nonumber
\end{eqnarray}
With such a subtraction we have a simple expression for $f_{sl} (g)$,
\begin{equation}
f_{sl} (g) = f(g) \, \ln \frac{2 \sqrt 2 }{g} + \sqrt{2} \, g \, \tilde S^{extra}_{1} (g) \, ,
\label{fextra}
\end{equation}
where $f(g)$ is the usual cusp anomalous dimension and $\tilde S^{extra}_{1} (g)$,
supposed to be free of logarithms, can be easily studied numerically (a plot of the complete function 
$f_{sl} (g)$ in the range $g \in [0,5]$ can be found in fig.~1 together with a magnification of the weak-coupling behaviour).
From a general point of view, it is likely to expect that $\sqrt{2} \, g \, \tilde{S}^{extra}_1 (g)$ can be expanded in powers of $g$ as follows,
\begin{eqnarray}
\sqrt{2} \, g \, \tilde S^{extra}_{1} (g) =k_1 \,  g   + k _0+ \frac{k_{-1}}{g}+ O(1/g^2), \ \ \ \ g \to \infty \, .
\end{eqnarray}
The usual numerical analysis \cite{BBKS}, including up to $L=30$ Bessel functions in the Neumann
expansion, gives the following best fit coefficients:
\begin{eqnarray}
k_1^{num} = - 2.828426 \pm 0.000001 ; \ \ \ \ \ k_0^{num} = 0.3238 \pm 0.0001 ; \ \ \ \ \ k_{-1}^{num} = -0.01194\pm 0.00015.
\end{eqnarray}
Inserting this expansion into (\ref{fextra}) we obtain
\begin{eqnarray}
f_{sl} (g) = 2 \sqrt{2} \,  g  \left[ \ln \frac{2 \sqrt 2 }{g} - c_1 -
\frac{3 \, \ln 2}{2 \sqrt{2} \pi \, g} \ln \frac{2 \sqrt 2 }{g}
+\frac{c_0}{ 2 \sqrt{2} \, \pi \, g} - \frac{\textrm{K}}{ 8 \pi^2 g^2} \ln \frac{2 \sqrt 2 }{g}+ \frac{k_{-1}}{2\sqrt{2}\,g^2}+ O(\frac{\ln g}{g^3}) \right] \, ,
\end{eqnarray}
where $\textrm{K}=\beta(2)$ is the Catalan's constant and $c_1=-k_1/(2 \sqrt 2)$, $c_0=k_0\pi$. Now, we may also infer the plausible analytic values\footnote{They seem also to be in agreement with the expansion by Freyhult and Zieme \cite {FZ} }
\begin{eqnarray}
c_1 = 1, \ \ \ \ \,  c_0 = 6 \ln 2 - \pi,\ \ \ \ \ \, \, k_{-1}= \frac{4 \textrm{K}-9 (\ln 2)^2}{4 \sqrt{2} \pi^2}= -0.0118253\dots \,\, .
\end{eqnarray}
In particular, for twist two this value of $c_0 = 6 \ln 2 - \pi$ entails the non-published evaluation $c= 6 \ln 2 + \pi$ by N. Gromov, reported in the added note of \cite{BFTT}.

Eventually,  for generic twist $L$ plugging into (\ref{subleading}) the asymptotic value of the first generalised scaling function $f^{(1)} (g)=-1+O(e^{-\frac{\pi g}{\sqrt{2}}})$ \cite{FGR1}, we end up with
\begin{equation}
f_{sl} (g, L) = 2 \sqrt{2} \,  g  \left[ \ln \frac{2 \sqrt 2 }{g} - c_1 -
\frac{3 \, \ln 2}{2 \sqrt{2} \pi \, g} \ln \frac{2 \sqrt 2 }{g}
+ \frac{c_0+(2-L)\pi}{2 \sqrt {2} \pi g}
 - \frac{\textrm{K}}{ 8 \pi^2 g^2} \ln \frac{2 \sqrt 2 }{g}+ \frac{k_{-1}}{2\sqrt{2}\,g^2}+ O(\frac{\ln g}{g^3}) \right]   \,\, .
\label {Efinal}
\end{equation}
This shows clearly how, in this asymptotic expansion, the only trace of the twist comes up in the piece $c_0+(2-L)\pi$ and thus cancels out completely, at order $O(s^0)$, in the
asymptotic (large $g$) expansion of  $\Delta -s=\gamma +L$. This cancellation is due to the simple asymptotic expansion $f^{(1)} (g)=-1+O(e^{-\frac{\pi g}{\sqrt{2}}})$ \cite{FGR1} and holds for any $L$. Furthermore, it makes possible that the constant term
in $\Delta -s=\gamma +L$ at order $O(s^0)$ be $\frac{c}{\pi}=\frac{ 6 \ln 2 + \pi}{\pi}$ for any
twist and thus future comparison with string theory (which does not
distinguish between null and small values of $L$ \cite{BFTT}).

\begin{figure}[ht]
\setlength{\unitlength}{1mm}
\begin{picture}(100,104)(-3,0)
\put(0,0){\includegraphics[width=0.95\linewidth]{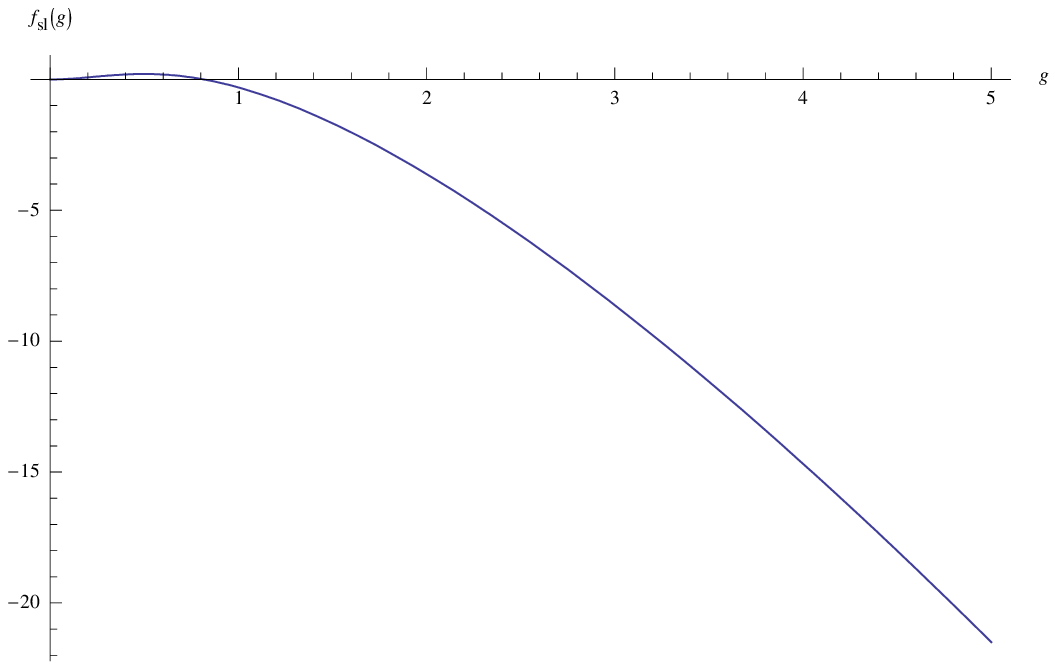}}
\put(139,40){\color{white}\rule{4.5mm}{40mm}}
\put(15,10){\includegraphics[width=70mm]{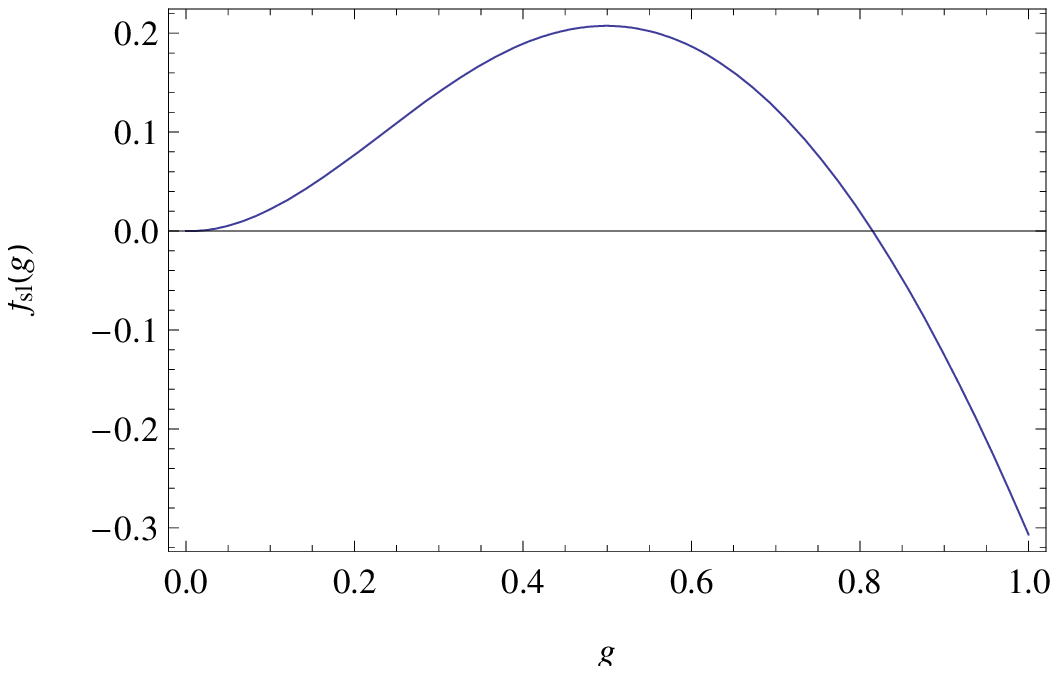}}
\put(10,87){\drawline(0,0)(32,0)(32,10)(0,10)(0,0)}
\put(10,87){\dottedline{3}(0,0)(13,-32)}
\put(42,87){\dottedline{3}(0,0)(39,-32)}
\end{picture}
\caption{\label{fig1} Plot of $f_{sl}(g)$ with magnification for $g\in (0,1)$.}
\end{figure}

\section{Summarising}
\setcounter{equation}{0}

In the present work we have performed a detailed study of the sub-leading (constant) contribution to the anomalous dimension of the twist operator in the large spin expansion.

To this aim, we fully exploited the results obtained in \cite{BFR} in order to write down a linear integral equation for $f_{sl}(g,L)$. This approach allowed us to study its weak-coupling expansion up to the order $g^{20}$ (and even further), and to access the strong coupling behaviour by means of a modification of well established numerical techniques. An entire range plot is depicted by Fig.~1.

The strong coupling regime is of particular interest, because it shows a leading logarithmic behaviour (with the 't Hooft coupling), that we are able to single out analytically by inspecting the structure of the inhomogeneous term in the linear integral equation (see appendix A).Moreover, relying on the numerical analysis we are in the position to disentangle the terms of the asymptotic expansion; in particular, we confirm the string result in \cite{BFTT}, additionally fix the constant $c$ for the twist $2$ case and understand how this constant is the same for any other twist. 

It is also important to stress that our computations, based on the asymptotic Bethe Ansatz, are in full agreement with previous findings
\cite{KLOV, B&KLRSV, FRS, BFTT, BJL, BeFo}.

Finally, our formalism also allows the treatment and can be seen as a particular case of the high spin and large twist case,
when the ratio between the logarithm of the spin and the twist is fixed: this is the subject of the contemporaneous paper \cite{FIR}.

\vspace {1.5cm}

{\bf Acknowledgements} D.F. ought to particularly thank M. Beccaria, D. Serban, I. Kostov, G. Korchemsky, A. Tseytlin, V. Forini, D.Volin and B. Basso for sharing their knowledge. Moreover, we thank D. Bombardelli, G. Infusino for discussions and suggestions. We acknowledge the INFN grant {\it Iniziative specifiche FI11} and {\it PI14}, the international agreement INFN-MEC-2008 and the italian University PRIN 2007JHLPEZ "Fisica Statistica dei Sistemi Fortemente Correlati all'Equilibrio e Fuori Equilibrio: Risultati Esatti e Metodi di Teoria dei Campi" for travel financial support. D.F. and P.G. acknowledge the Galileo Galilei Institute for Theoretical Physics as well as  M.R. the INFN/University of Bologna for hospitality. The work of P.G. is partially supported by the MCINN and FEDER (grant FPA2008-01838), the Spanish Consolider-Ingenio 2010 Programme CPAN (CSD2007-00042) and Xunta de Galicia (Conselleria de Educacion and grant PGIDIT06PXIB206185PR).

\vspace {1.5cm}

\appendix

\section{Strong coupling of the extra terms}
\setcounter{equation}{0}

We briefly analyse the strong coupling (asymptotic) behaviour of the
forcing terms of the system (\ref {Sextsystem}) for the extra coefficients
$S^{extra}_{r}(g)$.
From the asymptotic formula
\begin{equation}
\int _{0}^{\infty}
{dh}\, {h^s} \, \frac {J_{r}({\sqrt {2}}gh)}{e^h-1}=\sum _
{m=0}^{\infty}\frac {2^{m+s-1}B_m}{m!({\sqrt {2}}g)^{m-1}} \frac
{\Gamma \left (\frac {r+m+s}{2}\right )}
{\Gamma \left (1+\frac {r-s-m}{2} \right )} \, , \quad r+s \geq 1 \, ,
\label {intasy}
\end{equation}
where $B_n$ are the Bernoulli numbers,
we deduce the large $g$ behaviour of all the forcing terms in
(\ref {Sextsystem}), but the one involving $\tilde J_1$.
The strong coupling limit of such term
can be evaluated after writing it as
\begin{equation}
\int _{0}^{\infty}\frac {dx}{2}\frac {J_{2}(x)+J_0(x)-1}{e^{\frac {x}
{\sqrt {2} g}}-1} \, .
\end{equation}
Then we compute separately the strong coupling limits of the addends containing
$J_2$ and $J_0-1$, respectively. For the former, we use (\ref
{intasy}). For the latter, one uses the integral representation
\begin{equation}
J_0(x)-1=\frac {1}{2\pi}\int _{-\pi}^{\pi}d\theta
\left (e^{ix\sin \theta}-1\right )
\end{equation}
and then integrates first on $x$, getting an expression in terms of $\psi $ functions,
and finally on $\theta$, after developing the $\psi$ functions for large
argument.
With this procedure we get the asymptotic formula
\begin{equation}
g\rightarrow \infty \quad \Rightarrow \quad
\int _{0}^{\infty}\frac {dh}{h}\frac {\tilde J_{1}({\sqrt {2}}gh)}
{e^h-1}= - \frac {g}{\sqrt {2}}\ln \frac {g}{\sqrt {2}}+\frac {g}{ \sqrt {2}}
\left (\frac {1}{2} - \gamma _E \right )
-\frac {1}{2} + O \left (\frac {1}{g} \right ) \, .
\end{equation}
We see that such term at large $g$ behaves as $-\frac {g}{\sqrt {2}}\ln g$.
Therefore, it dominates the strong coupling behaviour of the forcing terms of
(\ref {Sextsystem}) .

\end{document}